\documentstyle[aps,preprint,amssymb]{revtex}
\begin{document}
\title{The coherent dynamics of photoexcited green fluorescent proteins} 
\author{Riccardo A. G. Cinelli, Valentina Tozzini, Vittorio Pellegrini, 
and Fabio Beltram} 
\address{Scuola Normale Superiore and Istituto Nazionale per la Fisica della 
Materia, Piazza dei Cavalieri 7, I-56126 Pisa, Italy} 
\author {Giulio Cerullo, Margherita Zavelani-Rossi, and Sandro De Silvestri}
\address{Istituto Nazionale per la Fisica della Materia and Centro di Elettronica 
Quantistica e Strumentazione Elettronica, Dipartimento di Fisica, Politecnico di 
Milano, Piazza Leonardo da Vinci 32, I-20133 Milano, Italy} 
\author {Mudit Tyagi and Mauro Giacca} 
\address{Molecular Medicine Laboratory, International Centre for Genetic 
Engineering and Biotechnology, Padriciano 99, I-34012 Trieste, Italy} 
\maketitle 
\begin{abstract}
The coherent dynamics of vibronic wave packets in the green fluorescent protein 
is reported. At room temperature the non-stationary dynamics following impulsive 
photoexcitation displays an oscillating optical transmissivity pattern with 
components at 67 fs (497 cm$^{-1}$) and 59 fs (593 cm$^{-1}$). Our results 
are complemented by {\it ab initio} calculations of the vibrational spectrum of 
the chromophore. This analysis shows the interplay between the dynamics of the 
aminoacidic structure and the electronic excitation in the primary optical 
events of green fluorescent proteins.

\end{abstract} 
\newpage 

The green fluorescent protein (GFP) of the {\it Aequorea victoria} jellyfish 
has emerged in recent years as a unique fluorescent label in several
biological studies\,\cite{gfplibro}. GFP is a large intrinsically-fluorescent 
protein (238 amino acids) characterized by a cylinder-shaped three-dimensional 
structure with a diameter of 24 \AA\ and a height of 42 \AA\,\cite{ormo}. 
The chromophore, located at the center of the cylinder, is a photoexcitable 
green-light emitter autocatalytically generated by the post-translational 
modification of a 3-amino-acid sequence (Ser65-Tyr66-Gly67)\,\cite{heim_pnas}.
It consists of the hydroxybenzyl side chain of Tyr66 (phenolic ring) and the 
imidazolidinone ring formed by cyclization of the tripeptide (heterocyclic ring).
The spectral characteristics of its emission and absorption bands have received 
a tremendous amount of attention\,\cite{ormo,heim_pnas,heim_nature,patterson} 
in an effort to understand GFP photophysics down to the single-molecule 
level\,\cite{pierce} and design mutants with optical properties tailored to 
specific needs. To date there exists a large set of GFP mutants with absorption
and emission bands ranging from the violet to the red part of the spectrum.
Recently time-dependent analyses have shed light on internal photoconversion 
mechanisms\,\cite{schwille_fsc,chattoraj}. However, a definitive microscopic 
model accounting for the optical properties of the chromophore and its blinking 
and photobleaching dynamics\,\cite{patterson,pierce,schwille_fsc} is still 
missing, partly due to a lack of precise information on the electronic states 
involved and to the limited effort in theoretical modelling. Only in recent years, 
in fact, a few electronic-structure calculations of the GFP chromophore in 
different protonation states have been performed\,\cite{vomi,wevo}.
 
Few-optical-cycle laser pulses allow ultrafast spectroscopy with unprecedented
temporal resolution and are opening new avenues in the study of optical properties 
of molecules\,\cite{rosker_prl}. Creation of non-stationary vibronic wave packets 
and their observation have already provided new insights into the microscopic 
mechanisms responsible for the optical activity of few simple physical and 
biophysical systems\,\cite{tokizaki_prl,nisoli_prl,zhu_prl,rhodopsin_science}. 
Observation of coherent phenomena in a protein, however, still remains a
challenge because of fast dephasing times. There exist only few reports on 
coherent dynamics in proteic systems whose functionality, however, is determined 
by non-proteic cofactors\,\cite{zhu_prl,rhodopsin_science}.

In this Letter we demonstrate that GFP ultrafast response after femtosecond 
laser excitation is dominated by the coherent dynamics of single-electron vibronic
wave packets, created in both the ground and excited states of the protein
chromophore\,\cite{single-electron}. The coherent dynamics manifests as an 
oscillatory modulation of the differential optical transmissivity. These results 
provide direct measurements of the chromophore collective vibrations during the 
optical process. In order to elucidate this dynamics, we also report the first 
{\it ab initio} calculations of the vibrational properties of the GFP chromophore 
based on the density functional theory (DFT).
 
The time-resolved technique here reported employed a pump-probe experimental
scheme\,\cite{nisoli_prl} with identical pump and probe pulses, resonant with 
GFP absorption. Sub-10 fs pulses centered around 500 nm were generated using 
a non-collinear optical parametric amplifier in the visible, pumped by the second 
harmonic of a Ti:sapphire laser\,\cite{giulio_apl}. The pulses had energies of 
about 40 nJ and spectral width allowing analysis in a broad wavelength range 
(490-550 nm). Samples consisted of GFP solutions at a concentration of 350 $\mu$M 
in volumes of 40 $\mu$l kept in a 0.5 mm-thick cuvette. Additionally, 
room-temperature steady-state absorption and emission spectra were measured.
In particular, absorption was measured in a 300 $\mu$l GFP sample at a 
concentration of 70 $\mu$M in a 1 cm-long cuvette. For these experiments we 
selected a representative GFP mutant, the enhanced GFP (EGFP), exhibiting a 
single absorption peak around 490 nm (associated to the anionic form of the 
chromophore\,\cite{brejc}) and improved brightness after blue-light excitation 
with respect to wild-type GFP\,\cite{patterson}. 

Figure 1 shows the steady-state absorption (dotted line) and emission (solid line) 
spectra of EGFP together with a schematic configuration-coordinate diagram for 
the anionic state. The absorption peak at 490 nm corresponds to the vertical 
electronic transition labeled as A. The emission lineshape displays two bands 
associated to transitions (B and C in the diagram) from the bottom of the excited 
state band into the first two vibrational levels of the ground state. The lower 
panel of Fig.\,1 shows the wavelength dependence of the average EGFP differential 
transmissivity ($\Delta$T/T) in the very first picoseconds following impulsive 
excitation. The observed positive values at the absorption and emission maxima 
can be assigned to ground-state absorption bleaching and stimulated emission from 
the excited state. In the inset of the lower panel of Fig.\,1, a typical time 
trace with low sampling frequency of $\Delta$T/T (at 500 nm) is reported. The 
signal remains constant during the first tens of picoseconds in agreement with 
expected values of fluorescence lifetime ($\approx$3.3 ns\,\cite{chattoraj}). 

Superimposed to the long-lasting value of the EGFP transmissivity we detected 
sub-picosecond oscillations during the first two picoseconds following excitation. 
In Fig.\,2 (left side) a representative time-dependent $\Delta$T/T at 530 nm is 
reported. Similar modulations were observed at other wavelengths. This oscillation 
pattern is a direct evidence of coherent dynamics of electronic wave packets
in the GFP chromophore and yields two distinct frequencies at 497 cm$^{-1}$ 
(period 67 fs) and 593 cm$^{-1}$ (period 59 fs) with dephasing time of about 1 ps. 
The two peaks in the power spectrum shown in the inset of the figure are clearly 
distinguishable and were found at unaltered position for the wavelengths studied. 
The measured frequencies correspond to the energy spacing between consecutive 
vibronic levels of the chromophore in the ground and excited states. In our case, 
for pump-pulse duration significantly shorter than the oscillation period, we 
expect the oscillatory signal to be dominated by excited-state dynamics, since 
the wave packet does not move greatly from the Franck-Condon region during 
excitation\,\cite{muka}. The ground-state vibrational frequency can also be 
roughly extracted from the energy spacing between bands B and C in the emission 
lineshape shown in Fig.\,1. This procedure gives a value of $\approx$660 cm$^{-1}$. 

However, in order to unambiguously establish that the dominant mode (at 
497 cm$^{-1}$) corresponds to the GFP dynamics in the excited state, we considered 
the spectral dependence of the amplitude and phase of the oscillatory 
pattern\,\cite{rhodopsin_science}. The coherent oscillation of a vibronic wave 
packet in the excited state, in fact, causes periodic wavelength shifts in 
stimulated emission. This should yield periodic changes in the intensity of the 
stimulated emission with maxima where the slope of the emission lineshape is 
large (at 500 and 530 nm, see Fig.\,1) and minima at the peak position and 
wings of the emission spectrum. In addition, the oscillatory pattern at 
wavelengths corresponding to the two sides of the emission peak must be out of 
phase. Similar arguments relate the vibronic wave packet in the ground state and 
absorption modulation to the absorption lineshape. The right side of Fig.\,2 
shows the amplitude (upper panel) and the relative phase (lower panel) of the 
measured transmissivity oscillations as a function of wavelength. This analysis 
reveals weak oscillations at wavelengths near the emission peak and at the wings, 
while larger on the steep sides. A strong phase change (approximately a phase 
inversion) centered around the wavelength of the emission maximum is observed. 
These data confirm that the origin of the dominant modulation at 497 cm$^{-1}$ is 
the vibrational dynamics in the excited state. We associate the other frequency 
to oscillations in the ground state in light of the results of the steady-state 
emission lineshape analysis and of our model-chromophore calculations (see below).

On the basis of these arguments, we are now able to elucidate the vibrational 
pattern driven by the optical excitation process and identify the coupling 
mechanism of the vibrational motion to the electronic excitation. To this end 
we performed electronic, structural, and vibrational calculations of the isolated 
EGFP chromophore in the anionic state within a DFT-based {\it ab initio} molecular 
dynamics approach\,\cite{capa}. The electronic structure was calculated by using 
a local density exchange and correlation functional with Becke and Perdew gradient 
corrections\,\cite{lda}. Soft first-principle pseudopotential\,\cite{vande} were 
used for the interactions between valence electrons and inner cores with a 25 Ryd 
energy cutoff for the plane-wave basis set. Simulations with 0.15 fs time-step
were performed in a 15 \AA\ cubic box, large enough to prevent interactions with 
the periodic images. 

In the top part of Fig.\,3 we show the HOMO (Highest Occupied Molecular Orbital) 
and LUMO (Lowest Unoccupied Molecular Orbital) of the chromophore. A charge 
transfer from the phenolic ring to the heterocyclic ring and a redistribution of 
the charge within each ring following electronic excitation is observed. This 
charge transfer is responsible for an increase of the proton affinity of the Tyr66 
heterocyclic-ring nitrogen and may be linked to the blinking 
dynamics\,\cite{vomi,vari}. 
 
The charge redistribution also causes changes in the strengths of some of the 
molecular bonds owing to the difference in bonding character between the ground 
and excited electronic states. This stimulates the dynamics seen in our 
experiments. The ground-state vibrational spectrum was calculated at T=300 K 
from the Fourier-transform of the velocity autocorrelation function on a 
$\approx$1.5 ps trajectory of Car-Parrinello molecular dynamics. The resulting 
spectrum is shown in the lower part of Fig.\,3 (dashed line). The vibrational 
frequencies in the region above 1000 cm$^{-1}$ agree within 5\% with recent Raman 
data on the EGFP chromophore\,\cite{behe} and correspond to (localized or 
collective) stretching modes\,\cite{vari}.

However, in thermally-equilibrated conditions not all the modes associated to 
the optical excitation have a significant spectral strength. We note that these 
modes are very unlikely to be observed in steady-state resonant Raman experiments
owing to fluorescence or sample degradation\,\cite{behe} but are accessible within 
our experimental approach. In order to specifically enhance these modes, we 
performed runs in appropriate non-thermally-equilibrated conditions reproducing 
the atomic displacements induced by the electronic excitation\,\cite{vari}. 
The corresponding spectrum is shown in Fig.\,3 (solid line). As expected, this 
simulation allows us to better identify the high-frequency streching modes in the 
range 1000-1650 cm$^{-1}$. Remarkably, the same simulation also yields vibrations 
below 800 cm$^{-1}$ corresponding to angular deformations of the chromophore. This 
fact provides evidence of intramolecular mode coupling among vibrations in two 
different and characteristic frequency ranges\,\cite{kidera}. The inset of Fig.\,3 
reports an enlarged view in the low-frequency region of the solid-line spectrum 
shown in the main panel. The arrow indicates the experimentally-measured 
ground-state frequency (593 cm$^{-1}$). At frequencies close to this one, we found 
two modes at 575 and 615 cm$^{-1}$. The identification of the specific one 
corresponding to the experiment is beyond our accuracy. However they both 
correspond to collective vibrations involving angular in-plane deformation of 
the rings (mainly the phenolic one) and of the bridge between them. We can 
therefore draw unambiguous conclusions on the relevant microscopic processes 
involved. 

The steps leading to the coupling of these low-frequency modes to the electronic 
photoexcitation can now be easily understood: in structures with a $\pi$-bonding 
system, the high-frequency stretching modes are usually directly coupled to the 
electronic excitation, since the primary effect of the induced electronic-density 
change is to shorten the single bonds and lengthen the double bonds (for example 
in retinals\,\cite{reti}). The double-ring structure of the GFP chromophore, 
however, introduces a strong geometric constraint that allows efficient coupling 
to the low-frequency angular modes responsible for the observed coherent dynamics. 
This process highlights the intramolecular coupling pathways and nuclear dynamics 
following photoexcitation and is a peculiar photophysical property of the GFP 
family.
 
In conclusion, we presented the coherent dynamics of single-electron vibronic 
wave packets following ultrafast excitation in EGFP. The analysis of coherent 
oscillations provided the vibrational frequencies of both the ground and excited 
states of the EGFP chromophore and allowed to evaluate the wave-packet dephasing 
time. The collective vibration excited during the optical process and its coupling 
mechanism to the electronic excitation have been identified by {\it ab initio} 
calculations. 
 
{\bf Acknowledgements}. One of us (V.T.) wishes to thank F. Buda for making 
available the code for Molecular Dynamics.

\begin{figure}
\caption{Upper panel: Enhanced green fluorescent protein (EGFP) absorption (dotted 
line) and fluorescence (solid line) spectra at room temperature. Letters indicate 
transitions between excited and ground electronic states as depicted in the 
schematic configuration-coordinate model. Lower panel: Wavelength dependence of 
the average EGFP differential transmissivity in the very first picoseconds after 
impulsive excitation (line is a guide to eye). Typical error bars and positions 
of the absorption and emission peaks are shown. The inset shows the differential 
transmissivity at 500 nm (acquired with low sampling frequency) as a function 
of delay between pump and probe pulses.} 
\end{figure} 
\begin{figure} 

\caption{Left side: Room-temperature enhanced green fluorescent protein (EGFP) 
differential transmissivity at 530 nm as a function of delay between pump and 
probe pulses. The inset shows the Fourier power spectrum of the data. Right side: 
Spectral dependence of the amplitude (upper panel) and relative phase (lower 
panel) of the measured transmissivity oscillations before damping (lines are 
guides to eye and typical error bars are shown).} 
\end{figure} 

\begin{figure}
\caption{Upper side: Simulated chromophore ($\rm C_{10}O_2N_2H_7^-$). The bonds 
with the protein backbone are cut at the level of the heterocyclic ring and 
saturated with hydrogen atoms. HOMO (Highest Occupied Molecular Orbital) and
LUMO (Lowest Unoccupied Molecular Orbital) are represented as isocharge surfaces. 
Carbon, hydrogen, and nitrogen are shown in black, white, and grey, respectively.
Lower side: Calculated vibrational spectra of the chromophore in 
thermally-equilibrated (dashed line) and non-thermally-equilibrated (solid line) 
conditions (see text). Inset: Enlarged view of non-thermally-equilibrated 
vibrational spectrum. The arrow corresponds to the experimentally-measured 
ground-state frequency (593 cm$^{-1}$).} 
\end{figure}


\begin{references}
\bibitem{single-electron}
The term ``single-electron wave packet'' is used to indicate that only one electron 
is photoexcited in the GFP molecules and that the following dynamics of each 
molecule is independent of the others.
\end{references}
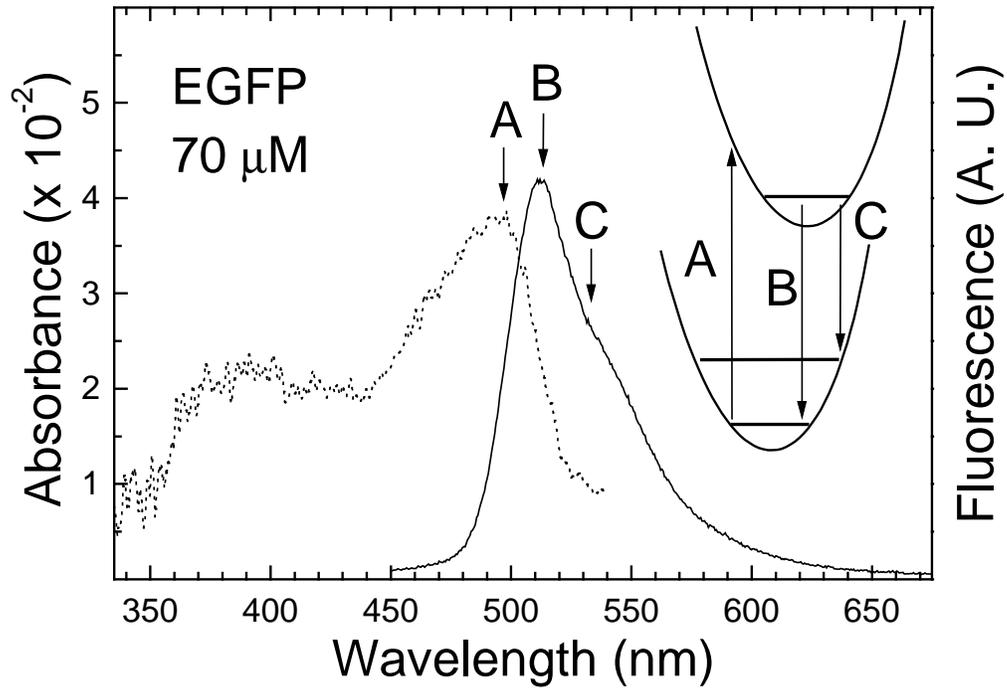
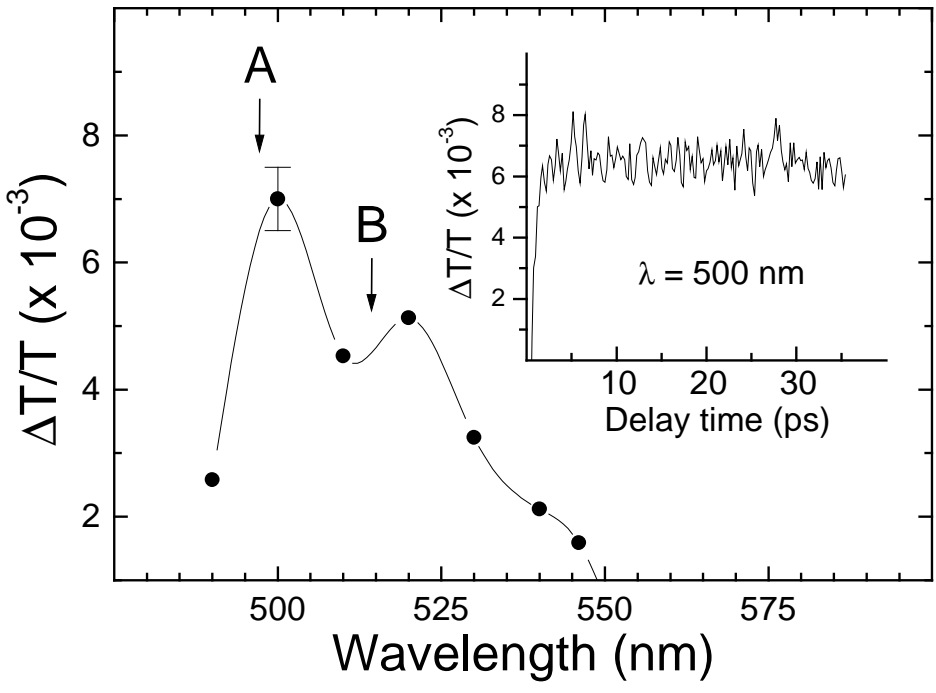
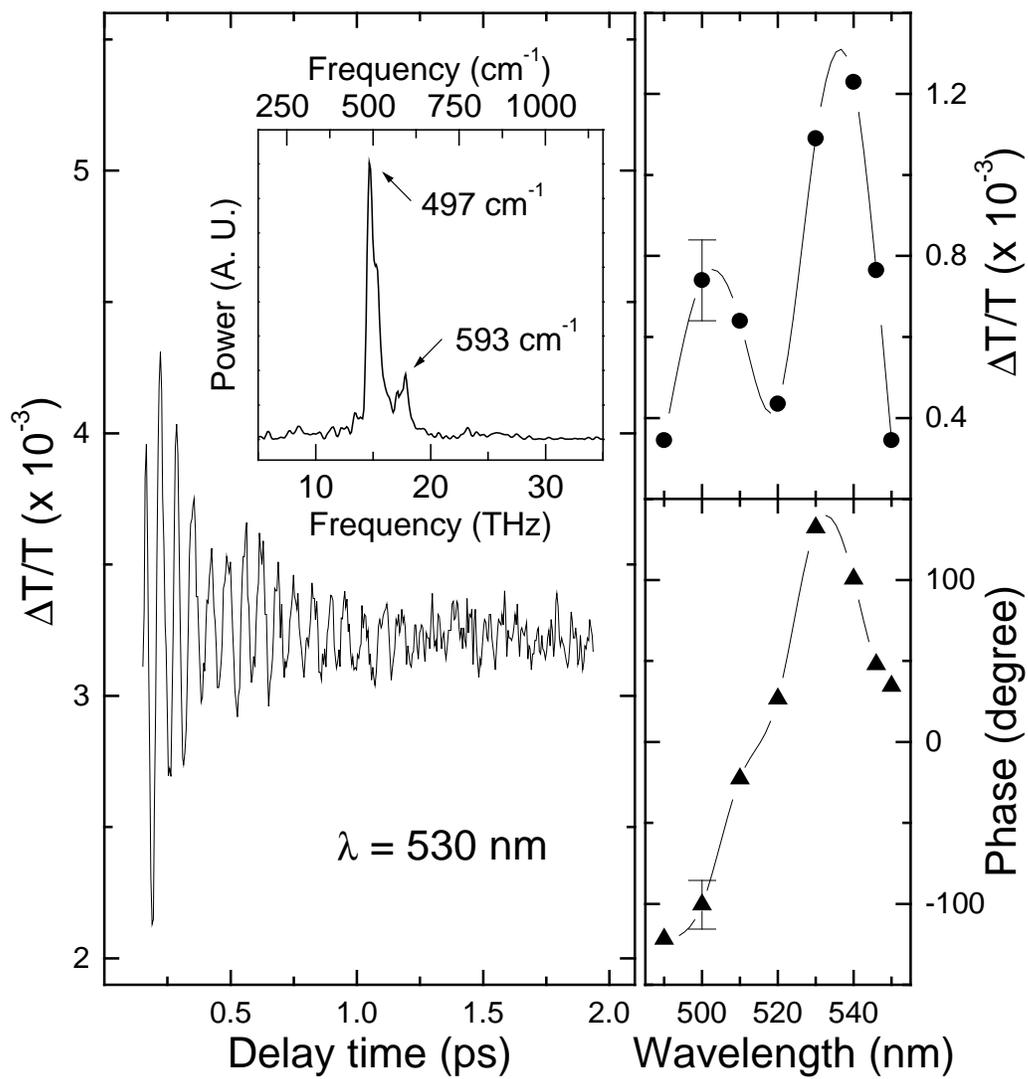
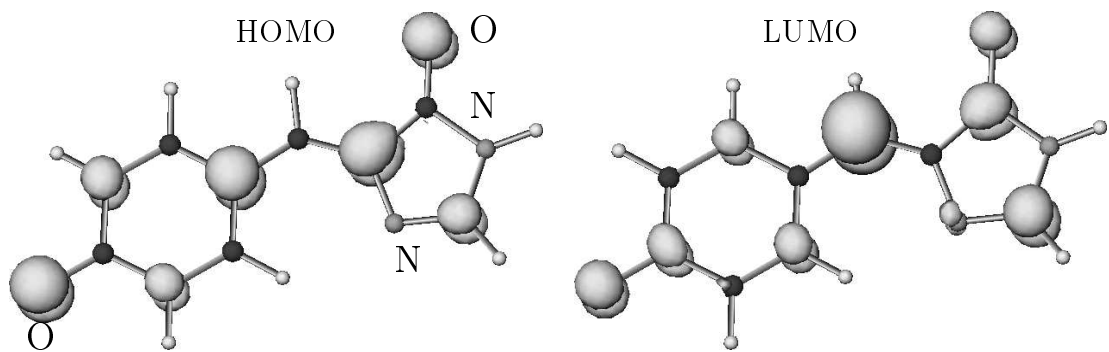
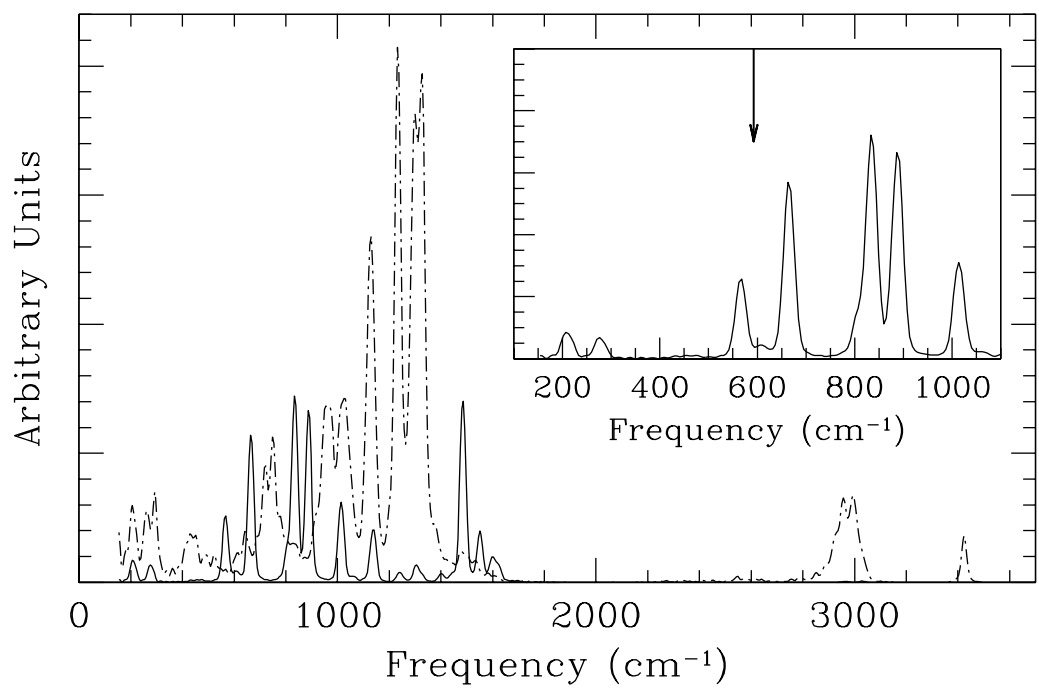
\end{document}